\documentclass[12pt]{iopart}

\usepackage{graphicx}
\usepackage{afterpage}
\begin{document}

\title{The role of stepped surfaces on the magnetotransport  in strained thin films of La$_{0.67}$Ca$_{0.33}$MnO$_3$}

\author{C. Beekman and J. Aarts.}
\address{Kamerlingh Onnes Laboratory, Leiden University, P.O. Box
9504, 2300 RA Leiden, the Netherlands}

\bibliographystyle{apsrev}
\date{\today}

\begin{abstract}\noindent
We report a comparative study of the properties of very thin films
of La$_{0.67}$Ca$_{0.33}$MnO$_3$ grown epitaxially under strain on
flat SrTiO$_3$ (STO) and on $1^{\circ}$-miscut STO. For the films
on flat STO the transport properties show well-known behavior,
namely values of the metal-insulator transition temperature which
are strongly reduced with respect to the bulk value. The reduction
in films on miscut STO is significantly less strong than in films
on flat STO, even though they appear similar as to strain state.
Moreover, the residual resistance at low temperatures is lower
than for the case of flat films. Magnetically, we find reduced
values of the saturation magnetization with respect to the bulk
value, indicating the presence of a dead layer in both cases. We
argue that the higher density of the step edges on the miscut
substrates lead to strain relaxation in the form of point defects
and an increased electronic bandwidth, which actually make the
electronic properties more robust.
\end{abstract}
 \maketitle

\section{Introduction}
Transition metal oxides with the perovskite structure are strongly
correlated electron systems which show diversity in physical
properties caused by the competition between charge, spin and
orbital degrees of freedom. Much work has been done in the last
decade to understand the underlying physics of the correlated and
semi-localized 3d electrons and their interaction with the
lattice. In particular for the manganites, showing the colossal
magnetoresistance effect (CMR) connected to the combined
metal-insulator (MI) and paramagnetic-to-ferromagnetic transition,
much progress has been made, as can be found in a number of
reviews \cite{tokura06,dagotto08,salamon01}. Still, even in bulk
materials the picture is still being refined, as was for instance
shown recently in the discovery of the existence of glassy
correlated phases in single crystals\cite{lynn07} and thin
films\cite{beekman5} of optimally doped
La$_{0.7}$Ca$_{0.3}$MnO$_3$. For thin films, a large amount of
work has gone into basic questions on their physical properties,
with the possibility of strain and interface engineering as issues
of special interest. In bulk manganites of the type
(RE$_{1-x})$A$_x$MnO$_3$ (RE is a 3+ Rare Earth ion, A is a 2+
alkaline ion) and at fixed RE to A ratio and therefore the
Mn$^{3+}$ to Mn$^{4+}$ ratio, the properties can be changed by
varying the radius of the 2+ ion\cite{machida4,tokura4}. This is
basically because the ion radius influences the structure of the
coupled network of MnO$_6$ octahedra, which changes the balance
between the itinerancy of the Mn 3$d$ electrons, and the strength
of the Jahn-Teller distortions which tend to trap electrons on the
Mn sites. In films this effect can be amplified by growing on a
substrate with a different lattice parameter, thereby putting the
film under tensile or compressive strain. For instance in films of
La$_{0.7}$Ca$_{0.3}$MnO$_3$ (LCMO; pseudocubic lattice parameter
$a_c$ = 0.386~nm) on SrTiO$_3$ (STO; $a_c$ = 0.391~nm) the
temperature of the metal-insulator transition T$_{MI}$ goes down
to 110~K for films with a thickness of around 10~nm, compared to a
bulk value of 260~K \cite{aarts98,doerr06,bibes01,yang04}. Below
about 5~nm the MI transition rather abruptly disappears, mainly
because the large strain leads to different crystal structures in
the film \cite{yang04}. At the same time, magnetic measurements
indicate the presence of a dead layer in the LCMO/STO
system\cite{deadlayer14,deadlayer24} of a few nm and nanoscale
phase separation \cite{bibes01}, and also in general a lowered
value of the saturation magnetization for larger thicknesses
\cite{aarts98}.

Gaining more understanding about these ultrathin films in the
regime around 10~nm is of interest, in particular since the highly
polarizable STO allows gating of the devices. In this regard, the
role of defects on the substrate surface, such as atomic steps, on
the electronic properties, have yet received very little
attention. Here we present a comparative study of LCMO films grown
on flat SrTiO$_3$ (STO; tensile strain) and on miscut STO
substrates, which shows that such steps can have a significant
influence, and are beneficial to several useful electronic
properties. The miscut substrates have an intentional
misorientation of the surface normal of 1$^{\circ}$ towards the
[010] direction, which leads to terraces of about 25 nm. We use
the miscut to investigate the sensitivity of film properties to
local variations of substrate surface and thus probing the effect
of local strain relaxation and disorder. From high resolution
transmission microscopy (HR-TEM) we find that basically all films
show the bulk $Pnma$ structure, with no clear differences between
the microstructures of films grown on flat and miscut STO. From
transport data we find that these film show the same trends in
$T_{MI}$ as found before, with a strong drop for films on flat
STO, and reaching a value of 110 K around a thickness of 10 nm.
For films on miscut STO, T$_{MI}$ stays significantly higher, not
coming below 140 K, but the low temperature resistivity in the
metallic state is found to be lower than for the films on flat
surfaces. We also discuss the issue of magnetization and show that
the saturation values for films in this thickness regime are often
lower than can be expected on the basis of a simple dead layer
picture. We distill the following physical picture. The high
density of the step edges leads to strain relaxation in the form
of point defects, and thereby to an increased electronic
bandwidth, with a higher T$_{MI}$ and  lower resistivity  The
presence of steps therefore, somewhat counter intuitively, has a
stabilizing effect on the electronic properties, and leads to more
effective charge transport.
\section{Experimental}
\begin{figure}[t]
\begin{center} \includegraphics[totalheight=0.35\textheight,
width=0.70\textwidth,viewport=10 10 450
400,clip]{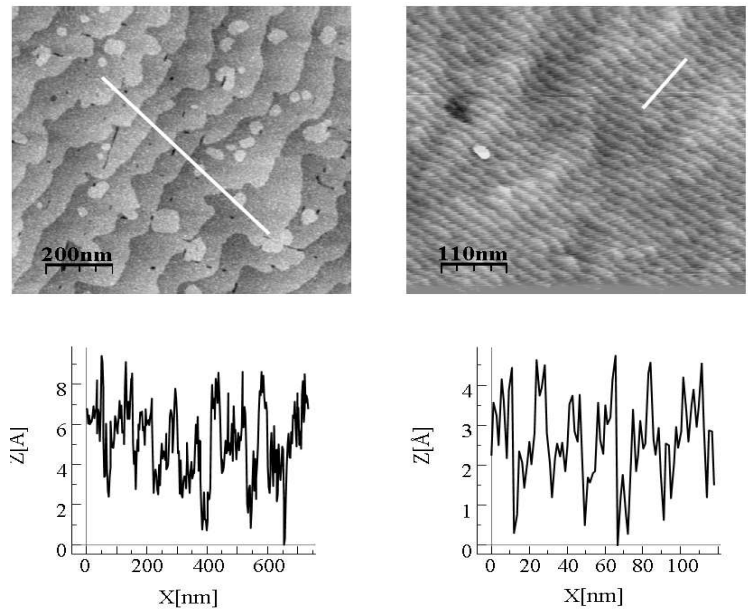}\end{center}
\caption{AFM images of LCMO films and corresponding profiles of (left) 15
nm LCMO on flat STO, L(15) (scale bar: 200 nm), and (right) 7 nm LCMO on STO with a
misorientation of 1$^{\circ}$ towards the [010] direction, L(7)$_{mis}$ (scale bar:
100 nm). }\label{fig1}
\end{figure}

\noindent
Epitaxial films of LCMO with thicknesses between 47 nm  and 6 nm were grown on (001)
STO substrates, by DC sputtering in oxygen pressure of 300 Pa at a growth
temperature of 840 $^{\circ}$C. The substrates have a misorientation of either
$<$0.2$^{\circ}$ in random direction, which we denote as flat STO, or 1$^{\circ}$
towards [010] direction. Here we define the nomenclature which we will use
throughout this paper to refer to our films. Films grown on flat STO are indicated
by L($d$), with $d$ the film thickness (rounded to the nearest integer value), and films
grown on misoriented STO by L($d$)$_{mis}$.
Before and after film growth we used Atomic Force Microscopy (AFM) to asses the
quality of the STO substrate and the LCMO film (see Fig.\ref{fig1}).
All films show clear unit-cell high step edges. The films grown on
flat STO show an average terrace length of ~75 nm and the films
grown on misoriented STO show an average terrace length of 20 nm,
identical to the terrace length of the substrate (see
Fig.\ref{fig1}). The thickness, growth rate and lattice parameters
of the LCMO films were determined by x-ray reflectivity (XRR) and
reciprocal space mapping (RSM) measurements,
respectively\cite{lat}. The average growth rate of our LCMO thin
films is 0.8 nm/min, which results in films with roughness of the
order of the dimensions of the unit cell.

\begin{figure}[htb!]
\begin{tabular}{c}\vspace{-5cm}
 \includegraphics[totalheight=0.35\textheight,
width=0.65\textwidth,viewport=10 10 250
200,clip]{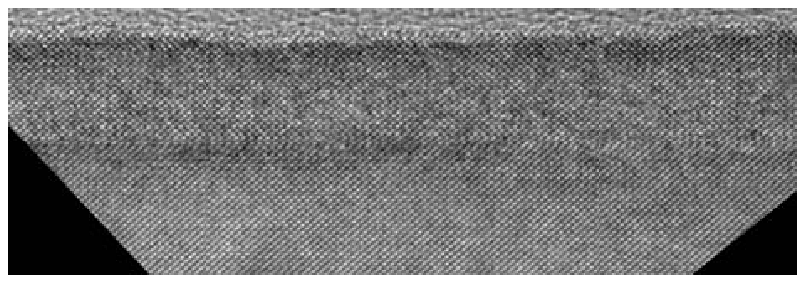}\\\vspace{-5cm}
 \includegraphics[totalheight=0.35\textheight,
width=0.65\textwidth,viewport=10 10 250
200,clip]{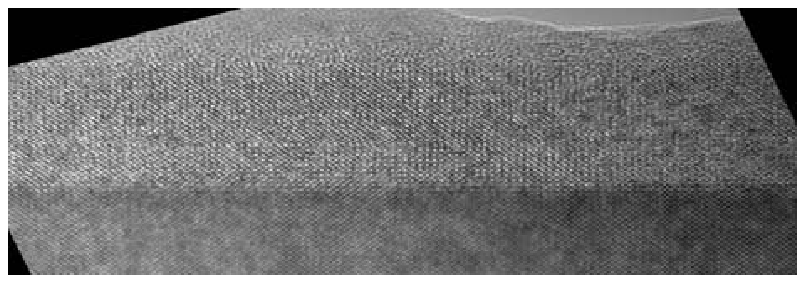}\\
 \includegraphics[totalheight=0.35\textheight,
width=0.65\textwidth,viewport=10 10 250
200,clip]{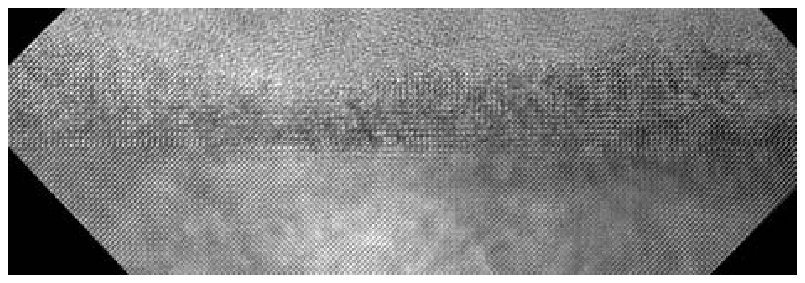}
\end{tabular}
\centering\caption{HR-TEM micrographs of LCMO films on STO, from top to bottom L(6),
L(10), L(7)$_{mis}$. The atom columns which are clearly visible in all cases set the
scale.}\label{fig2}
\end{figure}

We have characterized the microstructure using HR-TEM. The
specimens were prepared according to a standard cross-section preparation method.
Before insertion into the microscope, the specimens were plasma-cleaned for 1 minute
to prevent carbon contamination during the experiments. The analysis was performed
with a FEI TITAN equipped with a spherical aberration (C$s$) corrector and a High
Resolution Gatan Image Filter (HR-GIF) operated at 300kV. HR-TEM investigations on several specimens for each film confirm that our films
are epitaxial. The perovskite crystal structure of the films is close to cubic with
lattice parameter a$_c$ = 0.39 nm, but due to small rotations of the oxygen
octahedra it becomes orthorhombic (space group: $Pnma$). Using electron diffraction
we observed that throughout the films the bulk $Pnma$ structure is present with
lattice parameters of $\sqrt{2}a_c$, 2$a_c$ and $\sqrt{2}a_c$. For most films, the $b$ axis was found to be parallel to the interface normal (with
length 2$a_c$ in pseudocubic notation \cite{lat}). In HR-TEM mode (see
Fig.\ref{fig2}), and scanning along the interface, we did not observe any antiphase
boundaries or any domain type disorder, which is in line with previous reports
\cite{yang3a}. The amorphous layer visible at the top surface of the thin film in
Fig.\ref{fig2} (top image) is glue used during preparation of the sample. The HR-TEM
images shown in Fig.\ref{fig2} (samples L(6), L(10) and L(7)) all show a fully
epitaxial thin film.
%
There is a special reason to show a micrograph for sample
L(10), since this film has deviating properties compared to the typical film on flat
STO, as will be shown later. HR-TEM on this film showed no differences compared to the
other films. We also find that films grown on 1$^{\circ}$ misoriented STO
(Fig.\ref{fig2}c) are epitaxial with no clear influence of the step edges on the
microstructure of the LCMO film. We also do not observe any misfit dislocation at
the film-substrate interface in any of the films, which were investigated with
HR-TEM.

The magnetotransport properties were studied by measuring the
current ($I$) - voltage ($V$) characteristics as function of
temperature and in high magnetic fields. We used a PPMS (Physical
Properties Measurement System, Quantum Design) for temperature
control (T = 20 - 300 K) and for magnetic field control (H$_a$ = 0
- 9 T), in combination with an external current source and
nanovoltmeter. The films were patterned into
microbridges\cite{beekman} to determine the low temperature
resistivity. For the magnetization measurements we used an MPMS
(Magnetic Properties Measurement System, Quantum Design) with T =
10 - 300 K and H$_a$ = 0 - 5T.

\section{Transport properties}\label{transport}
\begin{figure}[t]
\centering\includegraphics[angle=0,height=9cm, width=12cm,angle=0]{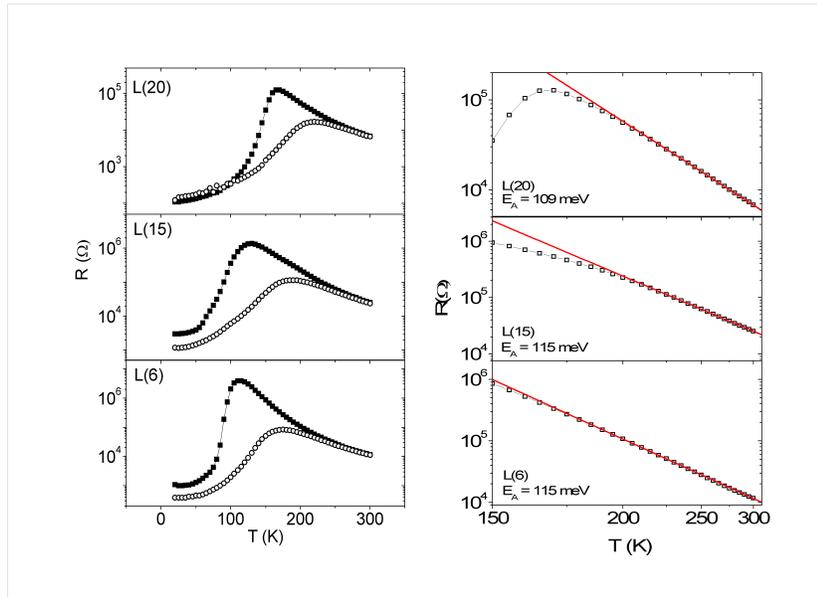}
\centering\caption{(Left) resistance $R$ vs. temperature $T$ for samples L(20) (top
panel), L(15) (center panel) and L(6) (bottom panel) as determined form the I-V
curves at an applied current of 0.1 $\mu$A. Squares: zero field; circles: H$_a$ = 5
T. (Right) log R vs. T in the temperature range 150 - 300 K for the same samples.
Note: the scale on the T-axis is reciprocal in order to show the 1/T-behavior. The
fit to extract the activation energy (E$_A$) of the polaron hopping process is also
shown (solid lines).}\label{fig7}
\end{figure}

We measured the $I$-$V$ characteristics in a temperature range of
20 - 300 K for films with varying thicknesses between 6 - 20 nm.
For films on both flat and miscut substrates the $I$-$V$ curves
were mostly linear, except in a small range around the transition,
where weak non-linearities were found. In Fig.\ref{fig7} we show
the temperature dependent resistance R(T) for three LCMO films
grown on flat STO, as determined at an applied current I of 0.1
$\mu$A.
The films show a clear metal-insulator transition accompanied by a resistance drop
of three orders of magnitude. All films show typical CMR effect, a reduction in
resistance of a few orders of magnitude upon application of a 5 T magnetic field.
For L(20) the transition temperature T$_{MI}$, which corresponds to the maximum
resistance value, is 170 K, which occurs approximately 100 K below T$_{MI}$ for bulk
LCMO. T$_{MI}$ is further reduced when the film becomes very thin, L(6) shows the
transition at T$_{MI}$ = 110 K, which is 60 K below T$_{MI}$ for L(20).
\begin{figure}[t]
\centering\includegraphics[width=8.06cm,height=6.4cm,angle=0]{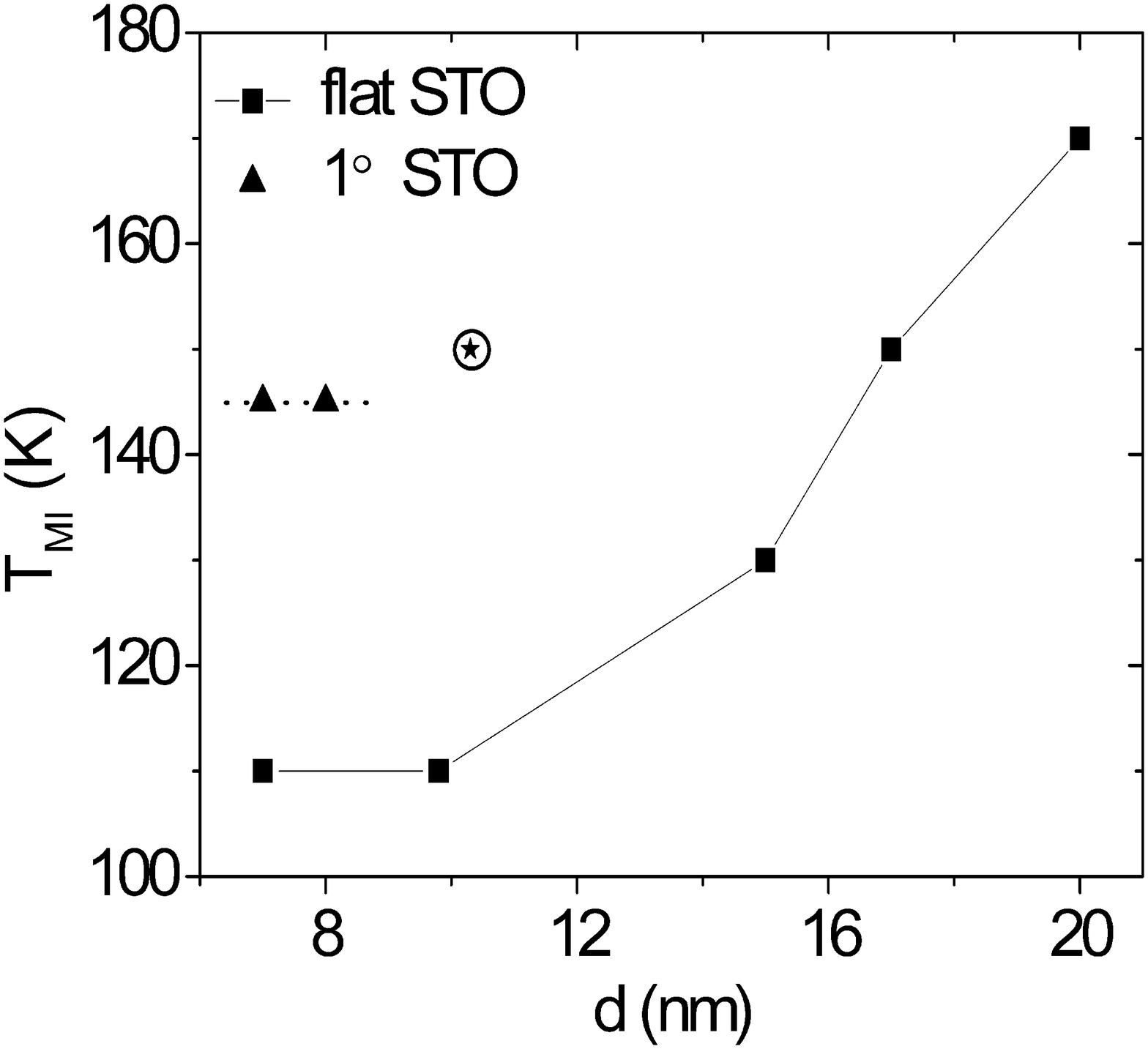}
\centering\caption{The dependence of T$_{MI}$ on film thickness
for LCMO films grown on flat (squares + star) STO and 1$^{\circ}$
misoriented STO (triangles). The star designates sample L(10).
Drawn and dashed lines are guides to the eye. }\label{fig8}
\end{figure}
The dependence on film thickness of T$_{MI}$ is shown in
Fig.\ref{fig8}. For films grown on flat STO the metal-insulator
transition is steadily shifted to lower temperature as the film
thickness is reduced. However, the film L(10) (indicated by
$\star$ in Fig.\ref{fig8}) grown on flat STO deviated from this
trend. There is a connected observation in
the magnetization measurements, which we will discuss below.

In the paramagnetic state, R(T) is expected to show activated
behavior. In Fig.\ref{fig7} (right panel), we plot log R vs. T
(with the T-axis reciprocal to show the 1/T-behavior). A linear
fit to the data provides the activation energy (E$_A$) for the
polaron hopping process. For most films the high temperature data
show activated behavior with E$_A$ = 110 - 120 meV (see Table
\ref{table}), independent of the film thickness, which is in the
accepted range of values for this temperature
regime\cite{palstra}. Deviations typically set in around 1.3x
T$_{MI}$ (for example see sample L(15) center panel
Fig.\ref{fig7}). In this respect at least, the thin films do not
behave different from bulk material. However, what happens in the
transition of such strained ultrathin films does not coincide with
the simple picture of variable range polaron hopping\cite{greaves,
JPCM}. Indeed, several studies have shown the formation of an
intervening phase consisting of glassy polaron
regions\cite{lynn07, beekman5} as the system is warmed through the
MI-transition. Moreover, we found earlier that the dimensions of
these correlated (static) polaron regions are sensitive to the
strain state of the films\cite{beekman5}.

\begin{figure}[t]
\centering\includegraphics[width=8.0cm,height=5.66cm,angle=0]{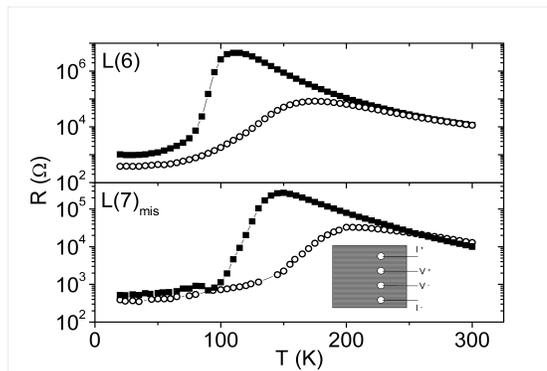}
\centering\caption{Resistance vs. temperature behavior of sample
L(6) (also shown in Fig. \ref{fig7}) and of L(7)$_{mis}$. For the
films on misoriented STO the transport properties were measured
with the current directed perpendicular to the step edges (see
inset for measurement geometry). The resistance values were
determined form the I-V curves at an applied current of 0.1
$\mu$A. The squares denote zero field; circles: H$_a$ = 5
T.}\label{fig9}
\end{figure}

A novel feature of our studies is the investigation of the effect
of unit-cell high step edges on the STO substrate surface on the
thin film properties. For these films the transport properties
were measured in a four-point geometry with the current directed
perpendicular to the step edges (see inset Fig.\ref{fig9}). Their
thickness was determined from HR-TEM micrographs and is 7 and 8 nm
(L(7)$_{mis}$ and L(8)$_{mis}$). R(T) for L(7)$_{mis}$ was
measured at $I$ = 0.1 $\mu$A and is shown in Fig. \ref{fig9}, with
R(T) of L(6) shown for comparison.
The observed T$_{MI}$ for both films on 1$^{\circ}$ STO is 145 K,
so T$_{MI}$ is much less reduced than for films on flat STO (see
Fig.\ref{fig8}, triangles). However, from the HR-TEM and RSM
characterization it is clear that the films on misoriented STO are
fully epitaxial across the entire film thickness. The lack of
reduction in T$_{MI}$ for films on misoriented STO is step-induced
but not due to the loss of epitaxial relation with the substrate.
It has been shown before that strain relaxation in these materials
may occur in the form of dislocations in the film \cite{lippmaa4}.
From HR-TEM we did not observe any dislocations in our thin films,
however, point defects should still be present and the amount is
possibly enhanced by the presence of the steps. Finally, we
mention the values of the residual resistivity (see Table
\ref{table}), as determined from structured samples. Resistivity
values for films on miscut substrates are clearly lower even
compared to the thickest films grown on flat substrates.

\section{Magnetic properties}\label{mag4}
\begin{figure}[t]
\begin{tabular} {cc}
\includegraphics[width=6cm]{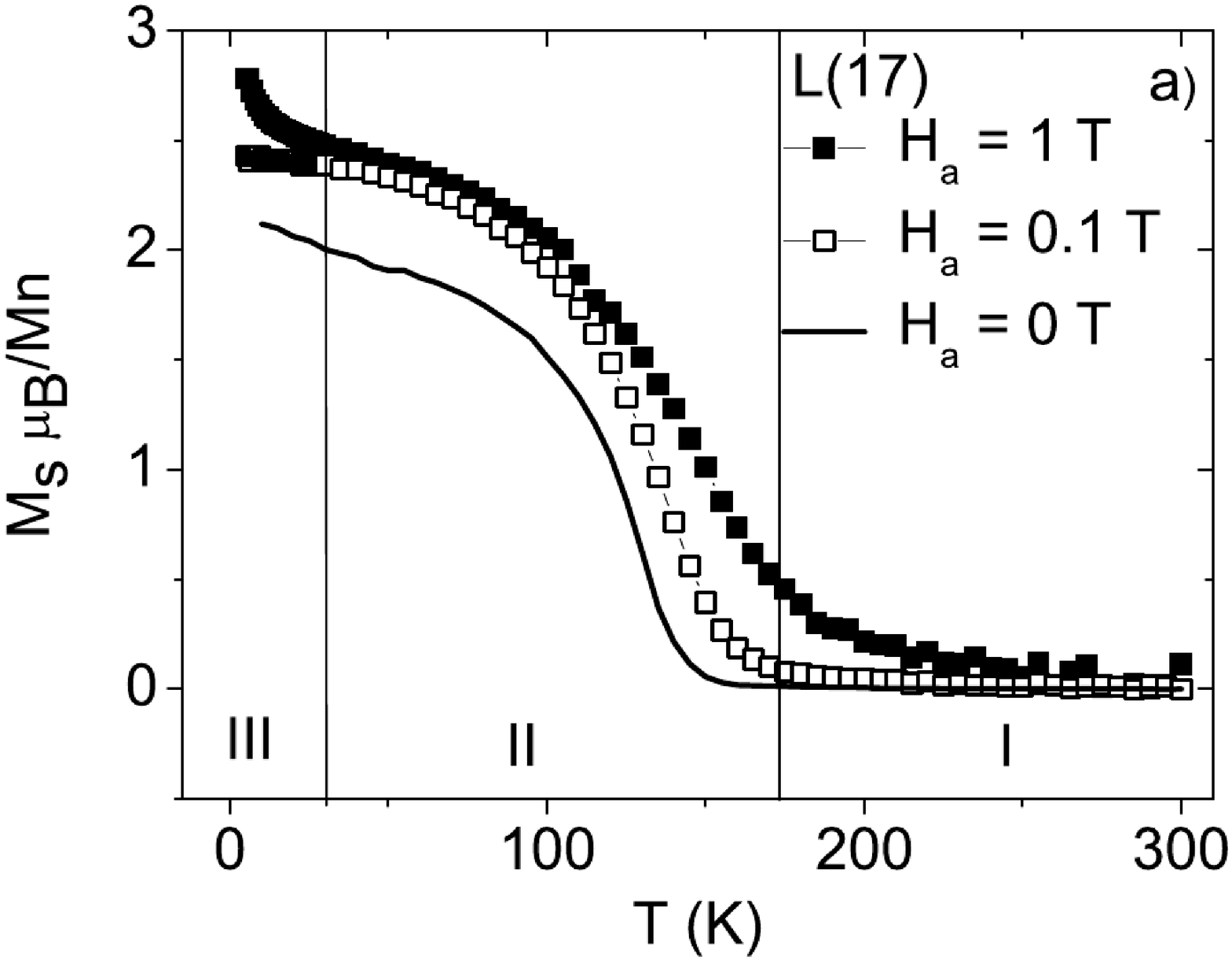} &
\includegraphics[width=6cm]{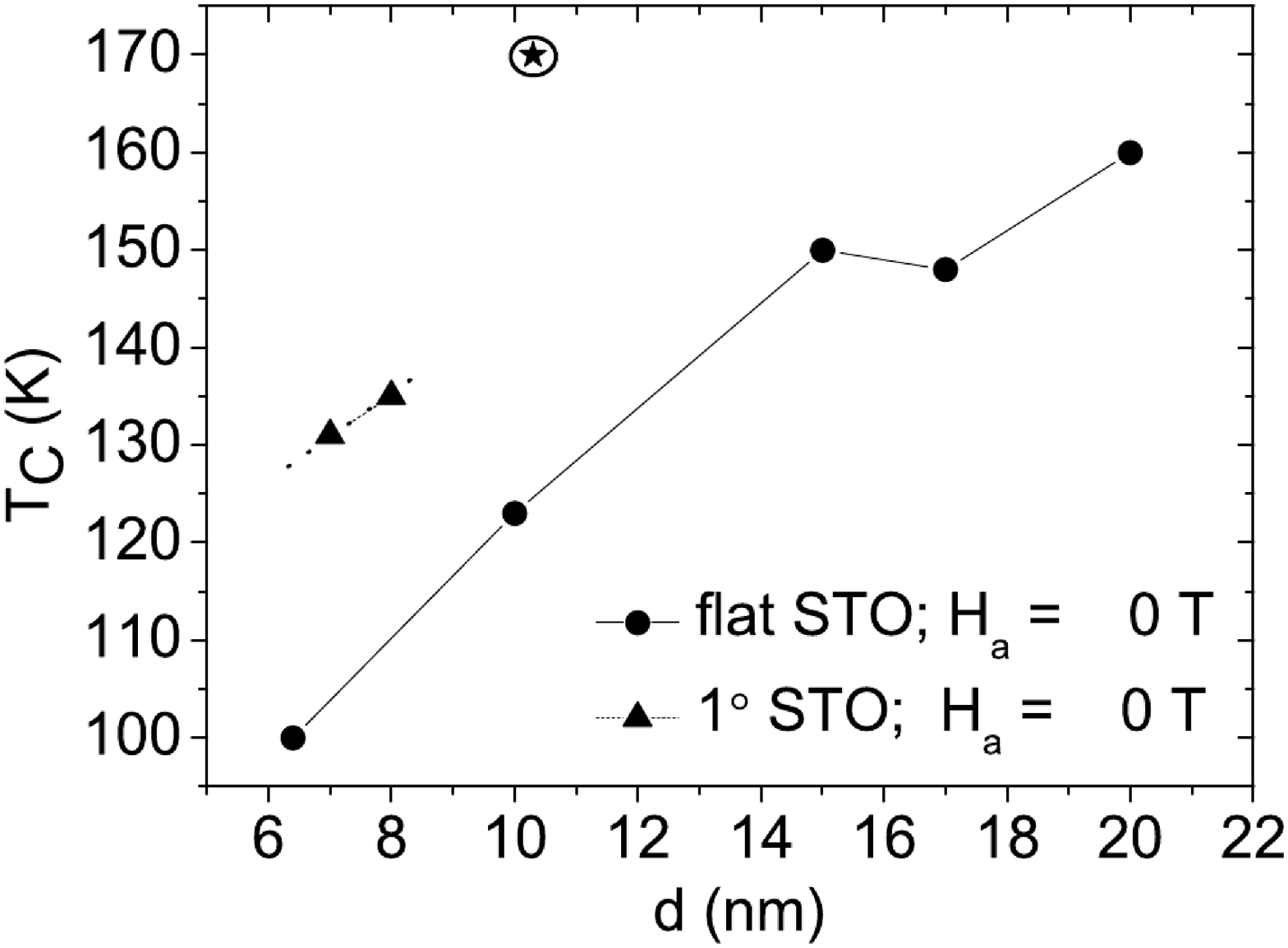} \\
\end{tabular}
\centering\caption{a) M vs. T behavior for L(17). The magnetization was measured in
H$_a$ = 1 T (closed squares), H$_a$ = 0.1 T (open squares) and H$_a$ = 0 T (line).
b) The Curie temperature as function of film thickness, determined from M vs T at
H$_a$ = 0 T (circles) and H$_a$ = 0.1 T (squares). The triangles show T$_C$ for two
films grown on 1$^{\circ}$ STO and the star shows T$_C$ for sample L(10). The drawn
and dashed lines are guides to the eye. }\label{fig10}
\end{figure}

\begin{figure}[hbt!]
\begin{tabular}{cc}
\includegraphics[width=6cm]{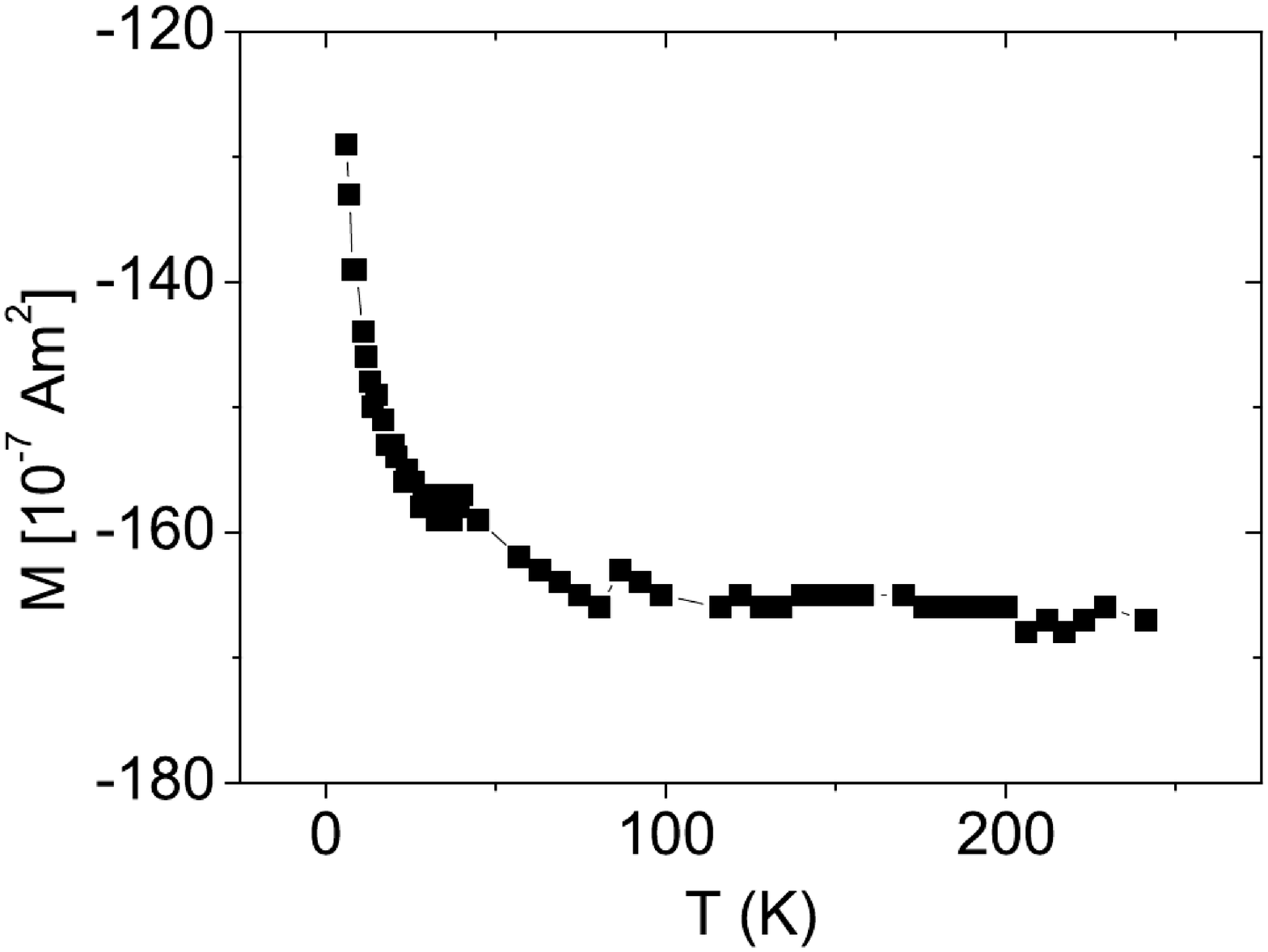} &
\includegraphics[width=6cm]{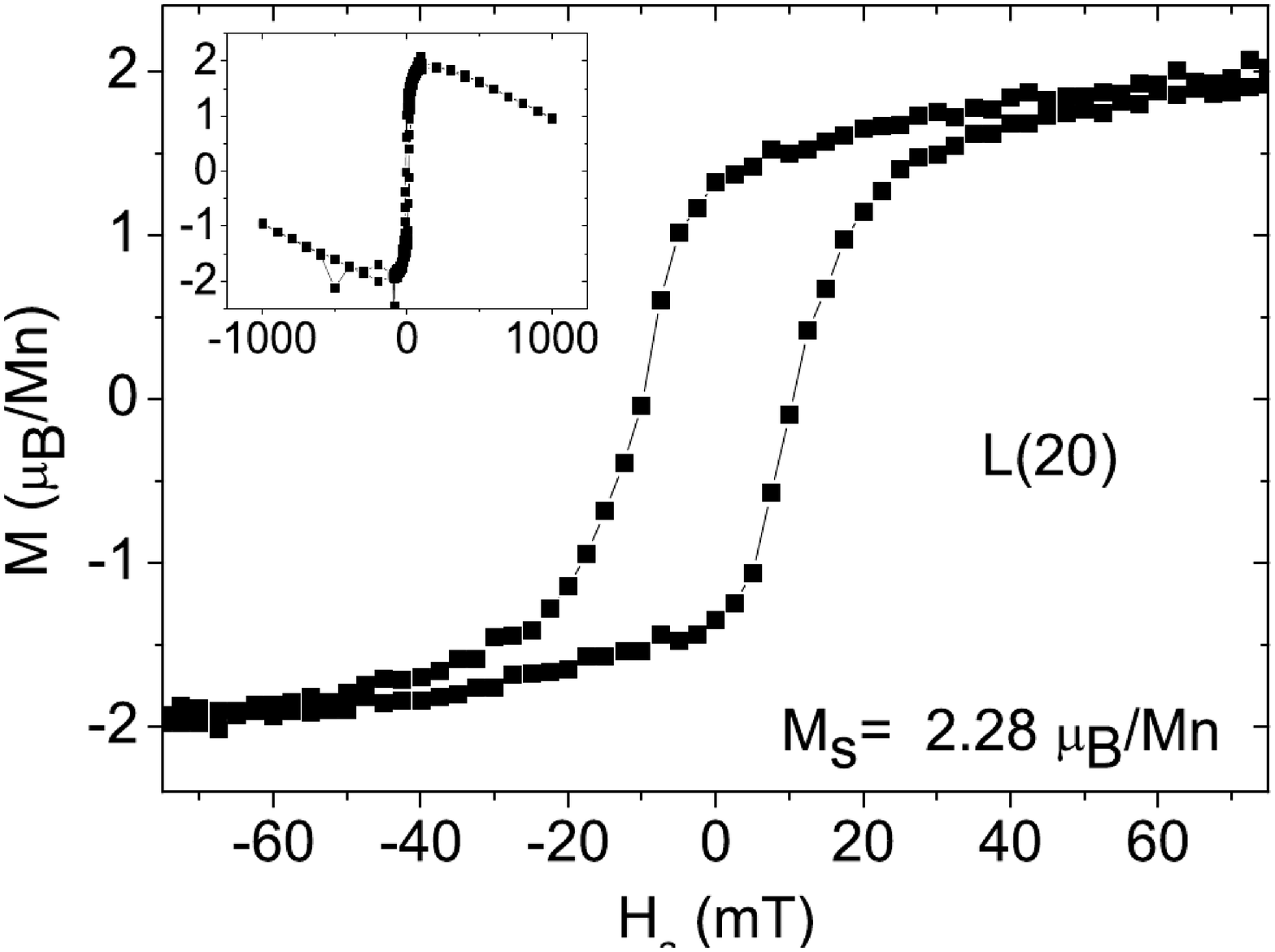} \\
\end{tabular}
\centering\caption{Left: Magnetization as function of temperature for a bare STO
substrate measured in H$_a$ = 1 T. Right: Magnetization as function of applied
magnetic field H$_a$ for the 20~nm thick LCMO film on flat STO. The saturation
magnetization $M_s$ at 1~T is 2.28~$\mu_B$ per Mn-ion. The inset shows M(H$_a$)
measured up to 1~T.}\label{fig11}
\end{figure}

\noindent Here we present the
magnetization behavior of the as-grown films on flat and misoriented STO substrates. Typical behavior of the magnetization $M$ vs. $T$ measured in
magnetic fields of H$_a$ = 0, 0.1~T, 1~T is shown in Fig.\ref{fig10}a for L(17). The
Curie temperature was determined from M vs. T, measured in zero magnetic field, by taking the intercept of the constant high temperature
magnetization with the linearly increasing M(T).
When the film thickness is reduced, we observe that T$_C$ is also shifted to lower
temperature as shown in Fig. \ref{fig10}b, and continues to coincide with the
temperature of the metal-insulator transition. For the measurement in H$_a$ = 1 T we assume that the
magnetization is saturated. In the M(T) behavior at 1~T (see Fig. \ref{fig10}a) the
magnetization shows a sudden increase below T = 30 K. The relative strength of this
upturn increases as the film thickness is reduced but the temperature at which the
upturn starts is constant. This feature is not an intrinsic feature of the LCMO thin
films. From Fig.\ref{fig11} it becomes clear that the $M$ vs. $T$ of a bare STO
substrate also shows an upturn below T = 30 K. Apparently, at low T a paramagnetic
contribution ($\chi$ = C/T, with C the Curie constant) dominates but disappears into
the diamagnetic background above T = 30 K. We surmise that the emergent
paramagnetism is due to the presence of impurities and/or defects in the bulk of the
substrate.
We have also investigated the field dependent magnetization at T = 10 K, shown for
L(20) in Fig.\ref{fig11}b. From these results we can extract the coercive field and
the value for the saturation magnetization (M$_s$) as function of film thickness.
The coercive field varies between +/-6 and +/- 15 mT for the different samples. The
magnetization is given in units of $\mu$$_B$ per Mn-ion and can be described, at
high fields, as M = M$_s$ + $\chi$H (see inset Fig.\ref{fig11}b).
The plotted data in the main graph is corrected for the
diamagnetic and paramagnetic contribution of the STO substrate by
subtracting $\chi \times H_a$ determined at high fields. The
values for M$_s$ are determined after correction by taking the
value for M at H$_a$ = 1 T. The theoretical saturation
magnetization for bulk LCMO would be 3.7 $\mu$$_B$ per Mn-ion. For
our films we observe slightly fluctuating values for M$_s$ but
always low compared to the expected bulk value. For L(20) in
Fig.\ref{fig11} the value is 2.28~$\mu_B$/Mn-ion. The lower values
indicate that a magnetically dead or weak layer is generally
present in our LCMO thin films. The only film showing almost
bulk-like saturation magnetization (i.e. no dead layer) is L(10).
This coincides with the enhanced value for T$_{MI}$ in this film.
Formation of the dead layer is probably sensitive to the initial
growth conditions, in particular to the oxygen stoichiometry of
the first layers. However, L(10) is the only film that deviates in
this respect, since films typically have the dead layer, this
makes it hard to investigate this effect further.

\section{Discussion}\noindent
\begin{table}[t]
\begin{tabular}{|c|c|c|c|c|c|c|c|c|c|}\hline
  Sample&\textit{d}& \textit{a$_{in}$} & \textit{a$_{out}$} & T$_{MI}$&T$_C$&  E$_A$& M$_s$& d$_{dead}$&$\rho$(T = 20 K) \\
     &nm& ($\dot{A}$) & ($\dot{A}$) & (K) & (K) &(meV) &   ($\mu$$_B$/Mn)& nm& [10$^3$ $\mu\Omega$cm] \\ \hline
   L(20) &20 &3.90&3.8113 & 170& 160 &109&   2.28& 7.6& 1.76 \\
   L(17) &17 &3.91&3.8135 & 150& 148 &114&   2.89 &3.7& 1.53\\
   L(15) &15 &3.90&3.8097&   130&150 &115&  1.96 &7.0& 22.6\\
   L(10) &10.3& 3.90&3.796 & 150& 170 & 114& 3.50 &0.5& 2.44\\
   L(6) &6.4& 3.91& 3.7979& 110& 100 &115 &  3.30 &0.7& 4.46\\
   L(8)$_{mis}$ &8 &-& -&  145 &135 & 119 &  2.41 &2.8& 0.80\\
   L(7)$_{mis}$ &7 &3.90& 3.7911 &  145 &131 & 116 & 2.22 &2.8& 1.23\\ \hline
\end{tabular}
\centering\caption{Summary of the measured values for the thin films presented in
this paper. Samples L(20)-L(6): on flat STO; L(8)$_{mis}$, L(7)$_{mis}$: on
1$^{\circ}$ miscut STO. The measured variables
are, \textit{d}: film thickness, \textit{a$_{in,out}$}: in-plane, out-of-plane
lattice parameters, T$_{MI}$: MI transition temperature, T$_C$: Curie temperature,
E$_A$: polaron hopping activation energy, M$_s$: saturation magnetization at T = 10
K at 1~T ($\mu_B$/Mn), the corresponding dead layer thickness
d$_{dead}$ and the resistivity as determined from tranport on microbridges patterned into the thin films\cite{beekman5}.}\label{table}
\end{table}
The main observations presented here are that for films on flat STO T$_{MI}$ (T$_C$) are reduced from the bulk value to a final value of 110 K for film thicknesses below 10 nm. This agrees nicely with the data presented by Bibes et al. \cite{bibes01} on similar strained films and confirms once more the effect of strain on T$_{MI}$. The novelty here is that fully strained films grown on miscut STO substrates exhibit significantly higher values for T$_{MI}$ compared to their flat STO counterparts. Transport properties of microbridges patterned into these films (reported elsewhere\cite{beekman5}) show that films grown on miscut substrates also have significantly reduced resistivity values (see table \ref{table}). Still, the previously reported strongly nonlinear behavior associated with the correlated polaron formation, as the device is warmed through the MI-transition, is present for all the films in this work. As the film thickness is reduced, the nonlinear behavior becomes more pronounced. Films on miscut STO show the nonlinear behavior in the onset of the transition, with a somewhat reduced intensity compared to film on flat STO with similar thickness.

All the above-mentioned observations in transport lead to the
following general picture. We believe that the presence of the
step edges lead to disorder (in the form of local
defects\cite{lippmaa4}). Perhaps counterintuitive, since Kumar et
al. \cite{kumar4} use the Holstein-Double Exchange model to show
that disorder enhancement of the polaron tendency in manganites
leads to increased resistivity and \textit{reduction} of T$_{MI}$.
Even more so since from the HR-TEM presented here there is no
visible evidence that the miscut alters the epitaxy in any way;
nor do we find any defects or dislocations. However, point defects
should still be present and could explain the enhanced itinerancy
of the miscut samples. Possibly, when the film tends to become
insulating the introduction of density of states inhomogeneities,
through defects and impurities, results in the formation of
additional conduction channels. This is in line with previous
reports \cite{pradhan5} which show that metallization of an
insulating phase of (La,Ca)MnO$_3$ is possible by introducing
valence variation on the Mn-site or by introducing defects in the
parent magnetic state. It seems probable that the introduction of
disorder also leads to disruption of the formation of the
correlated polaron regions, making it harder to form a homogeneous
glassy polaron phase.

Next we couple this to the magnetic properties of the films. For
all films the MI-transition coincides with T$_C$, which decreases
to a value around 100 K for thicknesses below 10 nm. The films
also show reduced values for the saturation magnetization,
indicating the presence of a dead layer. It is tempting to connect
the dead layer to reported NMR data\cite{bibes01}, which showed
that a non-ferromagnetic insulating (NFI) phase exists, at least
partly, in a region up to 5 nm from the interface. On the other
hand, our thinnest films show an almost full saturation
magnetization in 1~T. The films with thicknesses of 15~nm and
20~nm show a much lower saturation magnetization (even in 1~T)
than might be expected, leading to estimates of a dead layer
thickness which are not realistic. The reduction in magnetic
moment seems rather due to growth-induced defects, which are
antiferromagnetic in nature, but now not easily saturated. This is
a similar conclusion as was reached by us before \cite{aarts98},
but in contrast to e.g. Ref.\cite{bibes01}, and appears to depend
on the exact growth conditions of the films. Interestingly, the
films on miscut substrates, where relaxation has set in more
strongly, also show low values for $M_s$. More precisely, they
show the lower values found in thicker films grown on flat
substrates instead of the high values of the thinnest films. This
reinforces the argument that the step edges lead to disorder in
the in the form of local defects \cite{lippmaa4}, which then also
enhances T$_{MI}$.

\section{Conclusion}\noindent
In this paper we have presented a comparative study of the
magnetotransport properties of ultrathin LCMO films grown on flat
and on 1$^{\circ}$ miscut STO substrates.  For flat films, we find
the usual strain-induced reduction in T$_{MI}$ (T$_C$) as the film
thickness is reduced. However, for films grown on the miscut
substrates we find enhanced values for T$_{MI}$ (and T$_C$),
indicating that strain relaxation has set in more strongly due to
the presence of the step-edges. Together with a magnetically dead
layer, which is typically present for films on flat and miscut STO
alike, we deduce the following physical picture. The presence of
the step edges leads to strain relaxation in the form of point
defects, leading to an enhanced electronic band width due to
induced density of states inhomogeneities. The effect of step
edges is therefore somewhat different than might be expected.
Instead of adding to the disorder in such a way that the LCMO
films are driven towards an insulating state, they actually lead
to an electronically more stable situation.

\section{Acknowledgements}
We thank I. Komissarov for discussions, H. Zandbergen and M. Porcu
(National Centre for High Resolution Microscopy, Kavli Institute
for Nanoscience, Delft Technical University) for performing the
HR-TEM measurements and J.A. Boschker (University of Twente) for
performing the RSM measurements. This work was part of the
research program of the Stichting voor Fundamenteel Onderzoek der
Materie (FOM), which is financially supported by NWO.
\\


\begin{thebibliography}{99}
\bibitem{lat} The pseudocubic lattice parameters $a$ (in-plane) and $b$ (out-of-plane) as determined from
Reciprocal Space Mapping are shown in Table \ref{table}.
\end{thebibliography}
\end{document}